\documentclass[12pt,rmp,twocolumn,natbib,showkeys]{revtex4}

\usepackage{graphicx}
\usepackage{dcolumn}
\usepackage{bm}

\begin{document}

\title{Noise induced phenomena in point Josephson junctions}

\author{Anna V. Gordeeva$^1$,
Andrey L. Pankratov$^1$ and Bernardo Spagnolo$^2$}
\affiliation{$^1$
Institute for Physics of Microstructures of RAS, Nizhny Novgorod,
603950, Russia\footnote{e-mail: alp@ipm.sci-nnov.ru}\\
$^2$Dipartimento di Fisica e Tecnologie Relative, Universit\`a di Palermo and\\
CNISM-INFM, Unit\`a di Palermo, Group of Interdisciplinary
Physics\footnote{URL: http://gip.dft.unipa.it},\\
Viale delle Scienze, edificio 18, I-90128 Palermo\footnote{e-mail:
spagnolo@unipa.it}}

\date{\today}

\begin{abstract}
We present the analysis of the mean switching time and its standard
deviation of short overdamped Josephson junctions, driven by a
direct current and a periodic signal. The effect of noise enhanced
stability is investigated. It is shown that fluctuations may both
decrease and increase the switching time.
\end{abstract}
\date{\today}

\maketitle

\section{Introduction}

The investigation of nonlinear properties of Josephson junctions
(JJs) is very important due to their broad applications in logic
devices. Recently, a lot of attention was payed to Josephson logic
devices with high damping. In papers [Rylyakov $\&$ Likharev, 1999;
Bunyk \textit{et al.}, 2001] a description and analysis of the
entire system of single flux quantum logic elements are presented.
The noise properties of systems consisting of many elements may be
well understood from the noise properties of a single JJ. The
processes occurring in such devices are based on a reproduction of
quantum pulses due to $2\pi$ spasmodic change of the phase
difference of the overdamped JJs.

The use of high-$T_c$ JJs creates many new problems, such as the
thermally induced digital errors and the effect of spreading of the
switching speed due to thermal fluctuations. The analysis of
influence of thermal fluctuations on different characteristics of
JJs, such as the current-voltage characteristic or the life time of
superconductive state, has been made before on the basis of Langevin
approach [Barone $\&$ Patern$\grave{o}$, 1982; Likharev, 1986;
Ortlepp $\&$ Uhlmann, 2004; Ortlepp $\&$ Uhlmann, 2005]. Recently
experimental and theoretical work has been done on the mean
switching time or mean escape time from the metastable state in
periodically driven Josephson junctions [Yu $\&$ Han, 2003;
Pankratov $\&$ Spagnolo, 2004; Peltonen{\it et al.}, 2006; Sun {\it
et al.}, 2007].

\vspace{-0.1cm} The process of switching of a single JJ with high
damping under periodic driving was considered in [Pankratov $\&$
Spagnolo, 2004]. In that paper the influence of the resonant
activation (RA) and the noise enhanced stability (NES) phenomena in
the accumulation or suppression of timing errors in rapid single
flux quantum (RSFQ) devices was analyzed. Specifically an interval
of frequencies was found where the switching time increases with
increasing noise intensity. This effect is called noise enhanced
stability [Mantegna $\&$ Spagnolo, 1996; Agudov $\&$ Spagnolo, 2001;
Dubkov, Agudov $\&$ Spagnolo, 2004; Fiasconaro \emph{et al.}, 2005;
Spagnolo \emph{et al.}, 2007]. This paper presents the analysis of
the effects of thermal fluctuations on the switching time for a
single Josephson element with high damping, within the well-known
resistive model of a Josephson junction [Likharev, 1986]. The goal
of this work is to investigate in detail how the NES phenomenon
depends on the parameters of the periodical driving signal such as
the amplitude and the frequency.

\section{ General equations and statement of the problem}

The dynamics of a short overdamped JJ, under a current $i(t)$ is
given by the following Langevin equation

\begin{eqnarray}
\label{LE}
\omega^{-1}_c\frac{d\varphi(t)}{dt} =
-{\frac{du(\varphi )}{d\varphi }}-i_f(t),  \\
u(\varphi )= 1-\cos(\varphi )-i(t)\varphi, \quad \nonumber
\end{eqnarray}
where $\varphi$ is the phase difference of the order parameter
[Barone $\&$ Patern$\grave{o}$, 1982], $u(\varphi)$ is the
dimensionless potential profile (see Fig.~1.),
$\omega_c=2eR_NI_c/\hbar$ is the characteristic frequency of the JJ,
$I_c$ is the critical current, $R_N^{-1}=G_N$ is the normal
conductivity of the JJ, $e$ is the electron charge and $\hbar =
h/2\pi$, with $h$ the Planck constant. Here $i(t)=i_0 + f(t)$ is the
total current across the junction, $i_0$ is the constant bias
current, $f(t)$ is the driving signal, $i(t)=I(t)/I_c$,
$i_f(t)=I_f(t)/I_c$, and $I_f(t)$ is the random component of the
current. Because of thermal fluctuations, the random current may be
represented by white Gaussian noise
\begin{equation}
\label{2} \langle i_f(t) \rangle =0,\, \displaystyle{\langle
i_f(t)i_f(t+\tau) \rangle=\frac{2\gamma}{\omega_c} \delta(\tau )},
\end{equation}
where $\gamma =2ekT/\hbar I_c=I_T/I_c$ is the dimensionless
intensity of fluctuations, $T$ is the temperature and $k$ is the
Boltzmann constant.

Initially, the JJ is biased with a current across the junction
smaller than the critical one, that is $i_0 = (I_0/I_c) <1$, so as
the initial condition we take the location of the phase in a
potential minimum, $\varphi_0=\arcsin(i_0)$. A current signal
$f(t)$, such that $i(t) > 1$, switches therefore the junction into
the resistive state. In Fig.~1 we show the periodical potential
profile of the JJ and its extreme positions within which it varies
in time. The switching occurs not immediately, but at the later
time, which is called the switching time. Due to the noise the
switching time is a random quantity. We investigate therefore the
mean switching time (MST) $\tau$ and its standard deviation (SD)
$\sigma$. As a driving signal we choose a sinusoidal signal
$f(t)=A\sin(\omega t)$, where $\omega$ is the oscillation frequency
and $A$ is the signal amplitude.

According to the definition [Malakhov $\&$ Pankratov, 2002] the mean
switching time $\tau$ and its standard deviation $\sigma$ are
\begin{equation}
\begin{array}{ll}
\tau = \langle t \rangle = \int\limits_{0}^{\infty}tw(t)dt,  \quad
\langle t^2 \rangle = \int\limits_{0}^{\infty}t^2 w(t)dt, \\
\quad \quad \quad \quad \sigma = \sqrt{\langle t^2 \rangle-\langle t \rangle^2},
\end{array}
\end{equation}
where $w(t)=-\partial P(t)/\partial t$, $P(t)$ is the probability to find $\varphi$
within the interval $(-\pi,\pi)$.

\begin{figure}[h]
\begin{picture}(220,150)(10,10)
\centering\includegraphics[width=8cm,height=6cm]{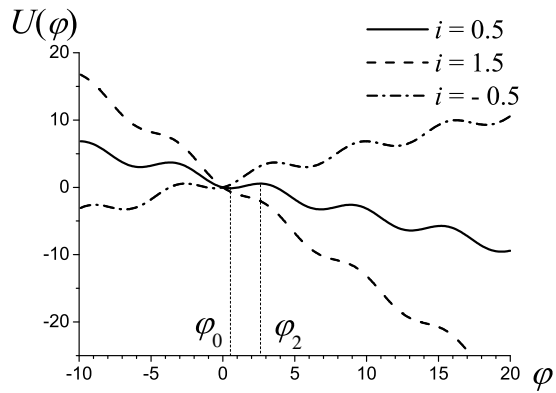}
\end{picture}
\parbox{230bp}{\caption{The potential profile of the JJ:
the extreme positions during the periodical variations ($i = -0.5$
and $i = 1.5$), and the intermediate configuration ($i = 0.5$). The
bias current and the signal amplitude are respectively: $i_0 = 0.5$
and $A=1$. The height of the barrier in the middle configuration of
the potential profile is approximately $0.7$.} \label{Fig1}}
\end{figure}

In Fig.~2 the MST is shown versus the signal frequency $\omega$ for
$i_0 = 0.5$, $A=1$ and for different values of the noise intensity,
namely $\gamma = 0.05, 0.2, 0.5, 1$. A frequency range, from $0.2$
to $0.4$, where the MST increases by increasing the noise intensity
is clearly visible. In all the figures the MST $\tau$ and the
frequency $\omega$ are normalized to $1/\omega_c$ and $\omega_c$,
respectively.

\begin{figure}[h]
\begin{picture}(220,150)(10,10)
\centering\includegraphics[width=8cm,height=6cm]{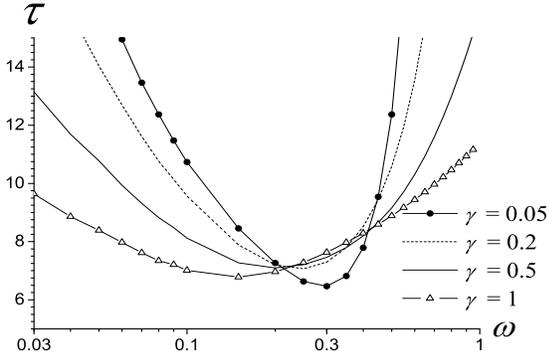}
\end{picture}
\parbox{230bp}{\caption{MST versus the signal frequency for different
values of the noise intensity: $\gamma = 0.05, 0.2, 0.5, 1$. Here
$i_0 = 0.5$ and $A=1$.} \label{Fig2}}
\end{figure}

In Fig.~3 we report the same quantity of Fig.~2, that is the MST
versus the signal frequency, for a smaller signal amplitude $A=0.7$
and different noise intensities, namely $\gamma = 0.005, 0.01, 0.05,
0.2, 0.5, 1$, and $i_0=0.5$. We see that, the frequency range where
the NES effect is observed is reduced, and it is shifted towards
smaller values of the driving frequency $\omega$, namely from $0.1$
to $0.2$.

\begin{figure}[h]
\begin{picture}(220,150)(10,10)
\centering\includegraphics[width=8cm,height=6cm]{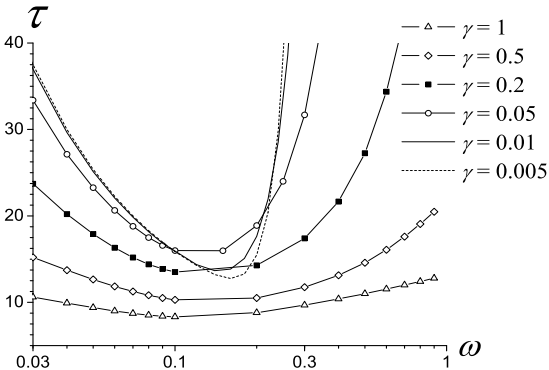}
\end{picture}
\parbox{230bp}{\caption{
MST versus the signal frequency for different values of the noise
intensity: $\gamma = 0.005, 0.01, 0.05, 0.2, 0.5, 1$. Here $i_0 =
0.5$ and $A=0.7$.} \label{Fig3}}
\end{figure}

In both Figs. 2 and 3, the MST has a minimum as a function of the
driving frequency. This is the signature of the resonant activation
phenomenon, investigated in a JJ in Refs. [Yu $\&$ Han, 2003;
Pankratov $\&$ Spagnolo, 2004; Sun {\it et al.}, 2007].

\section{ Results and discussion}

The enhancement of the switching time may be considered in detail by
plotting the dimensionless time MST as a function of the noise
intensity for different values of the signal frequencies. Different
behaviors of MST, depending on the values of the signal frequency,
occur for physical systems with metastable states [Agudov $\&$
Spagnolo, 2001; Dubkov, Agudov $\&$ Spagnolo, 2004]. Specifically a
nonmonotonic behavior and a monotonic one may be observed.

Let us start our consideration from small to large frequencies for
the case $i_0 = 0.5$ and $A=1$, see Fig.~2. For frequencies smaller
than 0.2 the character of the curves is similar: the MST decreases
monotonically with increasing noise intensity. This case is
presented in Fig.~4 by two values of frequency: $\omega = 0.05$ and
0.1. If the value of the noise intensity becomes greater or equal to
the height of the barrier, which is approximately equal to $0.7$ in
the middle configuration (see Fig.~1), the particle does not see the
fine structure of the potential. This is the reason why we restrict
to $1$ the values of the dimensionless noise intensity in the plots
(Figs. 2-9). So, we can see that the MST decreases, with respect to
the level of deterministic switching time (i.e. for $\gamma = 0$),
by increasing the noise intensity. The explanation of this behavior
is the following. For small noise intensity and very low
frequencies, the particle lies in the minimum for a long enough time
due to the slow variation of the potential profile. The switching
from the superconductive state to the resistive one does not occur
until the potential barriers disappear almost completely. By
increasing the noise intensity, the probability of the thermal
activated switching increases and as a result the MST decreases.

The next interval of frequencies which we consider is from 0.2 to,
approximately, 0.52, because for such frequencies the behavior of
the curves is similar. In this frequency range the NES effect
appears, the MST($\gamma$) - curve behaviour becomes nonmonotonic in
contrast with the previous case ($\omega < 0.2$). In Fig.~4 this
behavior is shown for three frequency values, namely $\omega = 0.5,
0.45$, and $0.4$. We note that in these three curves, after the
nonmonotonic behavior, the MST increases again with the noise
intensity $\gamma$ (for $\gamma > 1$). Due to the increasing values
of the signal frequency the escape process from the metastable state
becomes more rapid for low noise intensities and the MST for $\gamma
\rightarrow 0$ is lower with respect to the previous frequency range
($\omega < 0.2$). Due to the periodic variation of the potential
profile and for noise intensities smaller than the barrier height,
the particle starting from the minimum reaches a position near the
top of the barrier. Then there is an optimum range of the system
parameters, including the driving parameters, for which the particle
turns back at the metastable state because of the noise [Agudov $\&$
Spagnolo, 2001; Dubkov, Agudov $\&$ Spagnolo, 2004]. An enhancement
of the lifetime of the metastable state produces an enhancement of
the mean switching time. By increasing the noise intensity the
thermally activated escape increases too and the MST decreases. For
a further increase of the noise, for intensity values greater than
the barrier height ($\gamma > 1$), the potential profile
"\emph{seen}" by the particle is a linear one, which fluctuates
between the extreme positions of Fig.~1. The particle can reach more
easily positions near the state at $\varphi = 0$ and the MST,
therefore, restarts to increase again with the noise intensity.

\begin{figure}[h]
\begin{picture}(220,150)(10,10)
\centering\includegraphics[width=8cm,height=6cm]{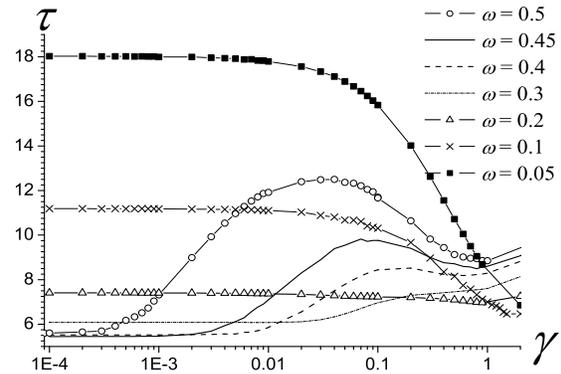}
\end{picture}
\parbox{230bp}{\caption{
MST versus noise intensity for seven different values of driving
frequencies, namely $\omega = 0.05, 0.1, 0.2, 0.3, 0.4, 0.45, 0.5$.
Here $A=1$ and $i_0=0.5$.}
\label{Fig4}}
\end{figure}

The nonmonotonic behavior shown in Fig.~4 may be explained also by
considering the competition of two factors in the above mentioned
frequency range: the influence of the returning force from the
lefthand side of the potential profile and the increase of the
variance of the phase position with increasing temperature. This
explanation is related to the static case when the switching event
starts because the value of the bias current is larger than the
critical one [Malakhov $\&$ Pankratov, 1996]. By increasing the
noise intensity the phase variance increases too. When the phase
variance is not so large, the main part of the phase probability
distribution is located on the flat part near the point
$\varphi=\arcsin(i_0)$, where the influence of the returning force
is low. Thus, the influence of variance dominates, and the mean
switching time increases with increasing of the noise intensity,
because the particle stays more near the metastable state. Further
increase of the phase variance, due to the noise, produces a larger
probability distribution of the particle, which reaches high slope
positions of the potential profile. Then the returning force begins
to affect from left to right because of the asymmetry induced by the
potential profile and does not permit, in this frequency range, the
probability distribution to move into the direction where the
potential profile goes up quickly. Therefore, the probability
distribution expands only into the right direction and the MST
decreases because the particle escapes. Finally, the noise intensity
becomes so large that it is possible for particle to move upstairs
along the potential. The returning force is not able to hold the
increasing of the variance anymore and the switching time increases
again.

By further increasing of the signal frequency ($\omega
> 0.5$, for the parameter values used to obtain the curves in Fig.~4)
a trapping phenomenon occurs. A threshold frequency $\omega_{th}$
exists which does not allow the particle to move to the next valley
during one period of the signal. This means that for driving
frequency $\omega > \omega_{th}$, the particle is trapped within one
period of the potential profile and, as a consequence, the MST
diverges (tends to an infinite value) without noise. The value of
the threshold frequency increases with increasing the bias current
and/or the maximal current across the junction [Agudov $\&$
Spagnolo, 2001; Dubkov, Agudov $\&$ Spagnolo, 2004]. The frequency
dependence of the switching time, for zero noise intensity, is
presented in the inset of Fig.~5. To be trapped the particle,
starting at $t =0$ from the position $\varphi_0$ (see Fig. 1) of the
potential profile in the middle configuration ($i = 0.5$), should
not reach the position at $\varphi_2$, in the lowest configuration
of the potential profile ($i = 1.5$), within a quarter of the signal
period. As a consequence the switching event from the
superconductive state to the resistive one doesn't occur. To
estimate the threshold frequency we calculate, in absence of noise,
the time $t$ spent by the particle to go from the point $\varphi_0$
to the point $\varphi_2$, in the most favorable case, when the
potential profile is fixed and is in the lowest configuration, with
the current $i = 1.5$ (see Fig.~1).

This time $t$, which is well approximated by one quarter of the
threshold period $T_{th}$, can be found from the solution of
Eq.~(\ref{LE}) without random current $i_f$ and for a constant value
of the total current $i(t)$ exceeding the critical one [Malakhov
$\&$ Pankratov, 1996]
\begin{eqnarray}
\label{AE} t=\frac{F(\varphi_2)-F(\varphi_0)}{\omega_c},
\quad \quad \quad \quad \quad \quad \quad \quad \nonumber \\
F(\varphi)=\frac{2}{\sqrt{i^2-1}}\arctan
\left(\frac{i\tan(\varphi/2)-1}{\sqrt{i^2-1}}\right).
\end{eqnarray}

From here the threshold frequency $\omega_{th} = 2\pi/T_{th} \simeq
\pi/2t$. The values of $\omega_{th}$ obtained from Eq.~(\ref{AE})
are marked by the vertical lines in the inset of Fig.~5.
\begin{figure}[h]
\begin{picture}(220,150)(10,10)
\centering\includegraphics[width=8cm,height=6cm]{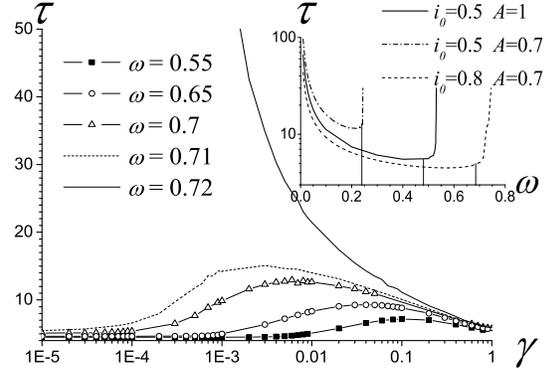}
\end{picture}
\parbox{230bp}{\caption{
MST versus noise intensity for different values of the driving
frequency, namely $\omega = 0.55, 0.65, 0.7, 0.71, 0.72$, which
approach the threshold frequency $\omega_{th} \simeq 0.72$ for a
bias current $i_0=0.8$ and a signal amplitude $A=0.7$. Inset: the
MST versus the driving frequency, from numerical simulations of
Eq.~(\ref{LE}) in the absence of noise ($\gamma = 0$), for three
groups of values of $A$ and $i_0$. The different values of
$\omega_{th}$, calculated by Eq.~(\ref{AE}), are marked by vertical
lines.} \label{Fig5}}
\end{figure}

Specifically we have: $\omega_{th} = 0.48$, for $i_0=0.5$ and $A=1$;
$\omega_{th} = 0.24$, for $i_0=0.5$ and $A=0.7$; $\omega_{th} =
0.69$, for $i_0=0.8$ and $A=0.7$. As it is seen from the inset of
Fig.~5, these values are very close to the exact values calculated
by numerical simulations of Eq.~(\ref{LE}) in the absence of noise
($i_f(t) = 0$). The most prominent NES effect is observed when the
frequency of the periodic signal is near the threshold frequency
(see Fig.~5). For $i_0=0.8$ and $A=0.7$ the maximum MST is
approximately three times greater than its value at $\gamma =
10^{-4}$.

In the next Fig.~6 the curves of MST are shown vs the noise
intensity for five frequency values near and larger than the
threshold $\omega_{th}$, that is for $\omega \geq 0.5 \approx
\omega_{th}$.
\begin{figure}[h]
\begin{picture}(220,150)(10,10)
\centering\includegraphics[width=8cm,height=6cm]{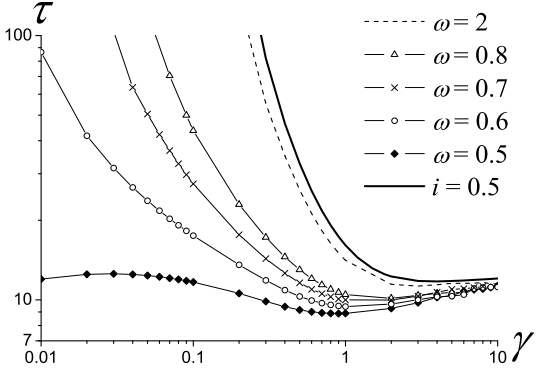}
\end{picture}
\parbox{230bp}{\caption{
MST versus noise intensity for five values of the driving frequency,
namely $\omega = 0.5, 0.6, 0.7, 0.8, 2$. The bold curve gives the
exact analytical behavior of MST for constant total current $i =
0.5$. Here $A=1$, $i_0=0.5$.} \label{Fig6}}
\end{figure}
Two distinct transient dynamical regimes are visible. For $\omega =
0.5$ we see a nonmonotonic behavior, with a finite value of $\tau$
for $\gamma \rightarrow 0$. For frequencies larger than the
threshold value $\omega_{th}$, all the curves are characterized by a
monotonic divergent behavior in the limit of small noise intensity
($\gamma \rightarrow 0$). In both dynamical regimes a minimum of MST
is present for a noise intensity value of the order of the barrier
height. By increasing the signal frequency the curves approach the
behavior of the MST obtained with a fixed potential profile,
corresponding to a constant total current $i = 0.5$. For high
frequency values, in fact, the fluctuations of the potential profile
are so rapid that the potential "\emph{seen}" by the particle is the
average potential, that is the middle configuration of Fig.~1, with
$i = 0.5$. In Fig.~6 the bold curve corresponds to the case of fixed
potential profile with constant total current $i=0.5$, and it was
calculated by using the exact analytical expression of MST, obtained
in [Malakhov $\&$ Pankratov, 1996].
\begin{figure}[h]
\begin{picture}(220,150)(10,10)
\centering\includegraphics[width=8cm,height=6cm]{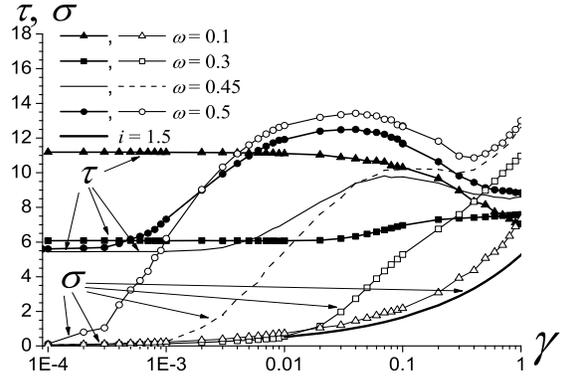}
\end{picture}
\parbox{230bp}{\caption{
MST and SD versus noise intensity for four driving frequencies,
namely $\omega = 0.1, 0.3, 0.45, 0.5$. The bold curve gives the
asymptotic analytical behavior ($\gamma << 1$) of SD for constant
total current $i = 1.5$. Here $A=1$, $i_0=0.5$.} \label{Fig7}}
\end{figure}

In the regime of frequencies above the threshold one the conditions
of stochastic resonance (SR) observation are fulfilled (see e.g.
[Borromeo $\&$ Marchesoni, 2000]). SR is manifested when the MST
does not exceed the period of signal and thus, the system response
becomes periodic owing to noise (see Fig.~5, curve $\omega = 0.72$
for which minimal MST is equal to $\approx 6$ and Fig.~6, curve
$\omega = 0.6$, minimal MST $\approx 10$).

In Fig.~7 the MST and its SD for four values of the signal frequency
are shown. We note that, rising from low to high frequencies near
the threshold $\omega_{th}$, monotonic and nonmonotonic behavior of
the MST correspond to the same behavior of the SD, respectively.
Moreover, in the limit of small noise intensity, all the SD
behaviors show lower values with respect to the corresponding MST
values, but with different behavior depending on how much the
frequency is close to the threshold $\omega_{th}$. For frequency
$\omega = 0.1$, in fact, the SD increases slowly in such a way that
it takes the same value of the MST for large noise intensity
($\gamma = 1$). But for greater frequency the noise intensity value,
for which the two curves of MST and SD cross, decreases.
Specifically we have the following cross point values: $\gamma =
0.04$ for $\omega = 0.45$ and $\gamma = 0.002$ for $\omega = 0.5$.
This means that when the driving frequency is close to the threshold
one the system becomes more unstable and even a small variation of
the noise intensity has a strong influence on the system. The curve,
shown in Fig. 7 by the bold line, is calculated for the case of the
total current $i=1.5$ and demonstrates a behavior of square root of
the noise intensity ($\sim \sqrt{\gamma}$). This curve was obtained
from the asymptotic analytical expansion of the SD in the small
noise limit $\gamma << 1$ and for a fixed potential ($\omega = 0$)
[Pankratov $\&$ Spagnolo, 2004; Gordeeva $\&$ Pankratov, 2006].
\begin{figure}[h]
\begin{picture}(220,150)(10,10)
\centering\includegraphics[width=8cm,height=6cm]{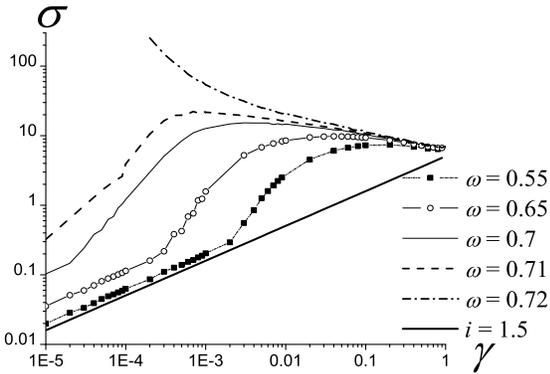}
\end{picture}
\parbox{230bp}{\caption{SD versus noise intensity for
five driving frequencies, namely $\omega = 0.55, 0.65, 0.7, 0.71,$
$0.72$. The bold straight line gives the asymptotic analytical
behavior ($\gamma << 1$) of SD for constant total current $i = 1.5$.
Here $A=0.7$, $i_0=0.8$.} \label{Fig8}}
\end{figure}
In the limit of $\omega \rightarrow 0$ the curves of the SD,
obtained by our simulations, approach the asymptotic curve obtained
for fixed potential. The behavior of the SD curve for $\omega =
0.1$, obtained by numerical simulations of Eq.~(\ref{LE}), is very
close to the theoretical asymptotic curve ($i=1.5$) not only in the
low noise limit but also at larger values of $\gamma$ ($\gamma = 0.1
- 1$). This feature of the SD($\gamma$) - curve behavior is shown in
more detail in the next logarithmic plot (Fig.~8), which
demonstrates the large rising of the SD for small noise intensities
with increasing frequency. In Fig.~8 the curves of the SD are shown
for the same frequency values and parameters used in Fig.~5 for the
MST behaviors. Again we note that the cross point between the MST
and SD curves, for frequencies close to the threshold one ($\omega
\simeq 0.72$ in this case), is at very low noise intensities. For
example for $\omega = 0.71$, the cross point is at $\gamma = 0.15
\cdot 10^{-3}$. After the threshold frequency, we recover for the SD
the same divergent behavior of the MST, but with larger values. This
fact confirms that the switching event in this region of parameters
is due to the noise. It should be noted also that for large noise
intensities ($\gamma>1$) all graphs of the MST and SD collapse to
the same curve, not depending on the signal frequency (see Fig.~5
and Fig.~8). This means that the new regime of switching is
unaffected by the signal and it is only due to the noise.

\begin{figure}[h]
\begin{picture}(220,150)(10,10)
\centering\includegraphics[width=8cm,height=6cm]{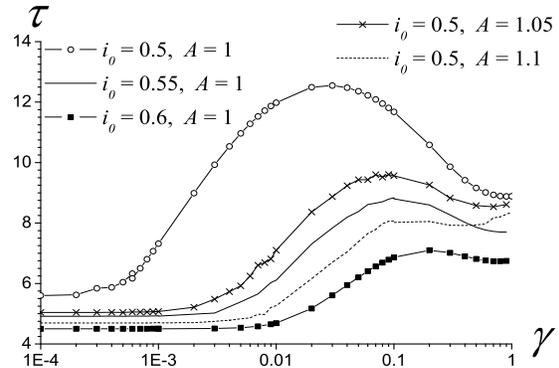}
\end{picture}
\parbox{230bp}{\caption{
MST versus noise intensity for two cases: (i) fixed bias current,
namely $i_0 = 0.5$ and $A = 1.05, 1.1$, and (ii) fixed signal
amplitude, namely $A = 1$ and $i_0 = 0.5, 0.55, 0.6$. Here $\omega =
0.5$.} \label{Fig9}}
\end{figure}

In Fig.~9 the curves of MST vs $\gamma$ for different values of the
bias current and signal amplitude are shown. We consider two cases:
(i) the bias current is fixed, namely $i_0 = 0.5$ and $A = 1.05,
1.1$, and (ii) the signal amplitude is fixed, namely $A = 1$ and
$i_0 = 0.5, 0.55, 0.6$. We choose $\omega = 0.5$ as a signal
frequency for which the NES effect is more pronounced. We see that
changes of the bias current modify the curves of MST more than
changes of the signal amplitude.

\section{Conclusions}

We analyzed the transient dynamics of a single overdamped Josephson
junction driven by a periodic signal. The conditions of existence of
the NES effect are investigated. We find that the enhancement of the
switching time is larger for frequencies of the periodic signal
close to the threshold frequency. In the region of the noise
intensity values in which the mean switching time begins to
increase, the standard deviation increases too but with different
behavior depending on the proximity of the signal frequency to the
threshold $\omega_{th}$. The interval of frequencies, in which the
NES effect is observed, depends on the maximal value of the total
current across the junction and on the value of the bias current.
There are two ways to change the value of the total current: by
changing the amplitude of the driving signal or the value of the
bias current. We find that the switching time changes more
significantly due to the bias current variation, than due to the
signal amplitude. We note that for small but nonzero values of the
McCumber-Stewart parameter the MST increases by several percent but
qualitative character of the curves remains the same, as in the
considered overdamped case.

\end{document}